\documentclass[fleqn,usenatbib]{mnras}

\usepackage{newtxtext,newtxmath,graphicx}
\usepackage{comment}

\usepackage[T1]{fontenc}

\DeclareRobustCommand{\VAN}[3]{#2}
\let\VANthebibliography\thebibliography
\def\thebibliography{\DeclareRobustCommand{\VAN}[3]{##3}\VANthebibliography}

\title[Weakness of soft state jets]{The stringent upper limit on jet power in the persistent soft state source 4U~1957+11}

\author[T.J. Maccarone et al.]{
Thomas J. Maccarone,$^{1}$\thanks{E-mail: thomas.maccarone@ttu.edu (TJM)}
Arlo Osler,$^{1,2}$
James C.A. Miller-Jones$^{3}$, 
\newauthor
P. Atri$^3$,
David M. Russell$^4$, 
David L. Meier$^5$, 
\newauthor Ian M. M$^{\rm c}$Hardy$^6$, and
Penelope A. Longa-Pe\~na$^7$
\\
$^{1}$Department of Physics \& Astronomy, Texas Tech University, Box 41051, Lubbock TX 79409-1051, USA\\
$^{2}$Department of Physics, Pima Community College, Tucson AZ, USA\\
$^3$International Centre for Radio Astronomy Research -- Curtin University, GPO Box U1987, Perth, WA 6845, Australia\\
$^4$Center for Astro, Particle and Planetary Physics, New York University Abu Dhabi, PO Box 129188, Abu Dhabi, UAE\\
$^5$Jet Propulsion Laboratory, Pasadena, CA, USA\\
$^6$ University of Southampton, Southampton\\
$^7$ Universidad de Antofagasta, Antofagasta, Chile\\
}

\date{Accepted XXX. Received YYY; in original form ZZZ}

\pubyear{2020}

\begin{document}
\label{firstpage}
\pagerange{\pageref{firstpage}--\pageref{lastpage}}
\maketitle

\begin{abstract}
We present extremely deep upper limits on the radio emission from 4U~1957+11, an X-ray binary that is generally believed to be a persistently accreting black hole that is almost always in the soft state.   We discuss a more comprehensive search for Type I bursts than in past work, revealing a stringent upper limit on the burst rate, bolstering the case for a black hole accretor.  The lack of detection of this source at the 1.07 $\mu$Jy/beam noise level indicates jet suppression that is stronger than expected even in the most extreme thin disk models for radio jet production -- the radio power here is 1500--3700 times lower than the extrapolation of the hard state radio/X-ray correlation, with the uncertainties depending primarily on the poorly constrained source distance.  We also discuss the location and velocity of the source and show that it must have either formed in the halo or with a strong asymmetric natal kick.
\end{abstract}


\section{Introduction}

Relativistic jets provide a strong impact on a variety of important
processes in the Universe.  The jets from supermassive black holes
provide one of the primary sources of feedback into the interstellar
medium \citep{SilkRees1998}, as well as one of the main mechanisms for accelerating the highest energy particles in the Universe \citep{Hillas1984}.  Because the jets may be powered by rotating black holes \citep{BZ}, understanding the mechanism for producing strong jets and the factors with which jet power correlates, may yield new means for probing the spin distribution of black holes. This may be especially useful for purposes like associating the environments of supermassive black holes with their spin histories, which requires large samples that probably preclude the source-by-source approaches such as reflection modelling \citep{Garcia2014} that have the best potential for giving precise spin estimates of individual systems. 

Direct studies of supermassive black holes suffer from a variety of challenges not applicable to stellar mass black holes.  Supermassive black holes show only weak, stochastic variability on human timescales, while stellar mass black holes often show factors of $10^6$ or more variability on timescales of months.  Furthermore, the
mass estimates for stellar mass black holes are generally more precise, and have systematics that are better understood, with clearer paths to improvement.  Thus, developing an understanding of how processes work in X-ray binaries and extrapolating those results to supermassive black holes is a fruitful approach (see e.g. \citealt{Merloni2003}).

One of the first clear results in understanding jet formation that came from X-ray binaries is that in spectral states dominated by soft X-rays, with strong thermal emission, the jet power is significantly weaker than in systems where the emission is dominated by hard X-rays \citep{Tananbum1972}.  Unfortunately, the excitement surrounding the discovery of the correlated X-ray and radio transitions from Cygnus~X-1 was, at the time, primarily that it provided evidence for a radio counterpart to the X-ray source, yielding good enough angular resolution to verify the correct optical counterpart.  It was thus about two decades before the significance of this discovery was fully appreciated, and it was found that the strong correlation between hard X-ray emission and radio emission was ubiquitous \citep{1995Natur.374..703H,Hannikainen1998,Fender1999}.

Theoretical work has since established a framework in which these correlations can be understood.  Jets are most likely to be powered by the poloidal components of the magnetic fields of their host accretion disks.  In the soft spectral states, the accretion disks are well modelled as geometrically thin, optically thick disks, with the spectra being produced as the sum of a series of diluted blackbodies \citep{ShakuraSunyaev,NovikovThorne,Davis2006}.  This approach yields a remarkably good spectral model for the existing data.  When the hard X-rays dominate, the emission is most likely to come from Compton upscattering in a geometrically thick, optically
thin medium \citep{ThornePrice,SunyaevTruemper}.  Since the geometric scale height of the accretion flow near the black hole is then dramatically larger, the poloidal component of the magnetic field should be larger.

Most black hole X-ray binaries spend most of their lives in the ``hard'' and quiescent spectral states. In typical soft X-ray transients, sources remain in faint hard states for long periods of time, undergo a disk instability which triggers a rapid rise in luminosity, and follow through a hysteresis loop in which they make transitions from hard to soft at higher luminosities than they make transitions from soft-to-hard \citep{Miyamoto1995,MaccaroneCoppi2003}.  Very strong radio flaring is generally seen in the hard-to-soft transitions, but not in the soft-to-hard transitions, and this has been interpreted as arising from the fast ejecta launched at the state transitions running into denser, slower moving ejecta from the previous long-lasting hard state when the hard-to-soft transition occurs, while, at the soft-to-hard transition, there are no nearby ejecta present (e.g. \citealt{Vadawale2003}).  In Cygnus X-3, the radio emission is enhanced in all spectral states relative to other sources, and the strongest flaring is seen at the return from the soft state to the more common hard states \citep{Koljonen2013}. This is explained best by a scenario in which the working surface for the jet in Cygnus X-3 is not other jet matter, but is the stellar wind, so that when the jet turns off for an extended period of time in the soft state, the wind has an opportunity to fill in the cavity that had been evacuated \citep{Koljonen2018}.  

Soft state detections from the stellar mass black holes that mostly are in hard states (e.g \citealt{Rushton2012}) may represent either {\it bona fide} soft state powering of the jet or leftover transient emission.  Soft state detections from Cyg X-3 are likely to be powered, at least in part, by the very strong free-free emission from the Wolf Rayet donor star's wind, or a collision between the WR wind and the disc wind from the accretor \citep{Waltman1996,Koljonen2018} rather than from the jet at all.   The ideal location to probe unambiguously the soft-state jet power is a low mass X-ray binary that is persistently or nearly persistently in a soft state.

One X-ray binary, 4U~1957+11, is a persistent source, nearly always in the high/soft state and is generally believed to have a black hole accretor (e.g. \citealt{Nowak2012,Gomez2015}, although see also \citealt{Bayless2011}).  While the persistent brightness makes it difficult to make precise mass and distance estimates, the persistent softness still makes it the ideal source in which to look for the most extreme reduction in radio power relative to standard correlations.  In \citet{2011ApJ...739L..19R}, we have already established this source to have its jet power suppressed by a factor of 330 to 810 relative to hard state X-ray binaries at the same X-ray luminosity with the uncertainties mostly due to distance uncertainties.  Here, we present the results of observations with a flux density 3.4 times lower, in conjunction with a slightly higher X-ray flux, providing evidence for jet suppression by a factor of at least 1500.  We also discuss its astrometric properties and the strong evidence that it either formed in the Galactic halo or formed with a very strong natal kick, an issue already raised in \citet{Nowak2012} which we can address in more detail here.

\section{Data}
\subsection{Radio data}
The Karl G. Jansky Very Large Array (VLA)  observed 4U~1957+11 for a total of 21 hours in 14 observing sessions between 20 February 2014 and 23 March 2014, with 17 hours on source, and 4 hours for calibration and slewing. The data were collected between 8 and 10 GHz under project code 14A-256.  Two of the observations (on 22 and 23 February) were very badly affected by radio frequency intereference in the nonlinear regime and were excluded.  The phase calibrator for the observations was J1950+0807, and the flux calibrator was 3C48.  The data were reduced using standard procedures in CASA \citep{2007ASPC..376..127M}, and we obtain a noise level of 1.07 $\mu$Jy/beam for the full data set.  

\subsection{X-ray data}
Because of the large number of dynamically scheduled radio observations, along with the high brightness of the source in X-rays, we chose to use all-sky monitor data in the X-ray band rather than to obtain new X-ray data.  The Monitor of All-sky X-ray Image (MAXI -- \citealt{MAXI2009}) data for the source show a weighted mean count rate of 0.103 cts\,s$^{-1}$\,cm$^{-2}$ in the date range from MJD 56712.5 to 56739.5, over which the data were taken.  This corresponds to a flux of $1.1\times10^{-9}$ erg\,s$^{-1}$\,cm$^{-2}$ (2-20 keV) for thermal models within the range typically seen from this source \citep{Nowak2012}. 

Additionally, we look at the X-ray data from proposal 50128 (PI: Nowak), the longest campaign on the source by the Rossi X-ray Timing Explorer \citep{Swank2006}.  This campaign had about 300 kiloseconds of data on source.  We look at these data to determine whether any Type I X-ray bursts took place.  The least frequent burster among the "banana state" neutron star systems is Ser~X-1 \citep{Galloway2008}, which bursts about once per 8 hours.  It had already been found in 26 kiloseconds of data that 4U~1957+11 shows no Type~I bursts \citep{Wijnands2002}, which already provided suggestive evidence that it is not a soft-state accreting neutron star.  With this additional data set of 300 kiloseconds, about 10 bursts would have been expected if the source bursted as frequently as Ser~X-1, and the lack of bursts thus provides very strong, albeit non-dynamical, evidence against a neutron star accretor.  

\section{The positional and binary system parameters for 4U~1957+11}
\begin{figure*}
\centering
\includegraphics[width=0.45\textwidth]{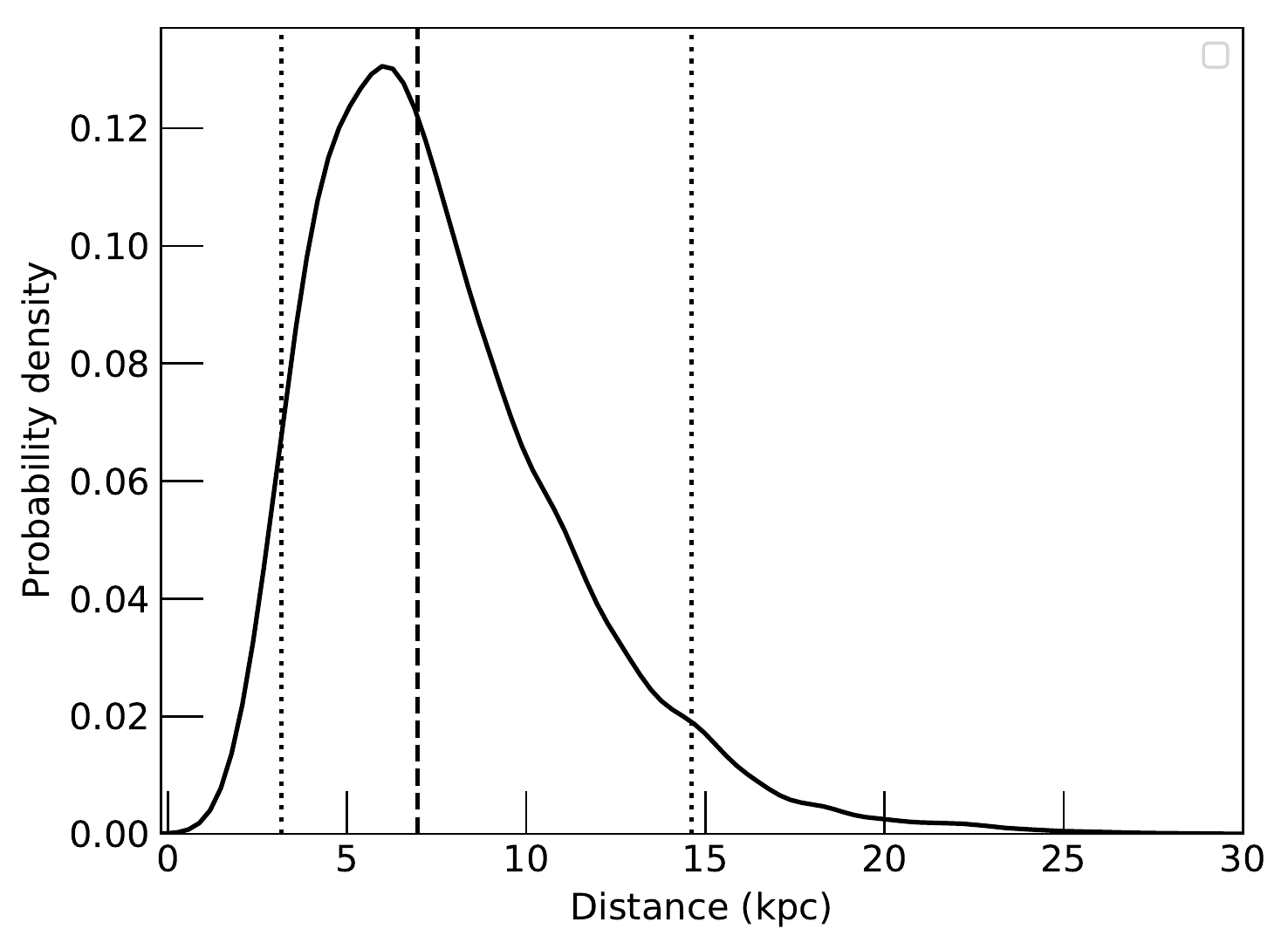}
\includegraphics[width=0.45\textwidth]{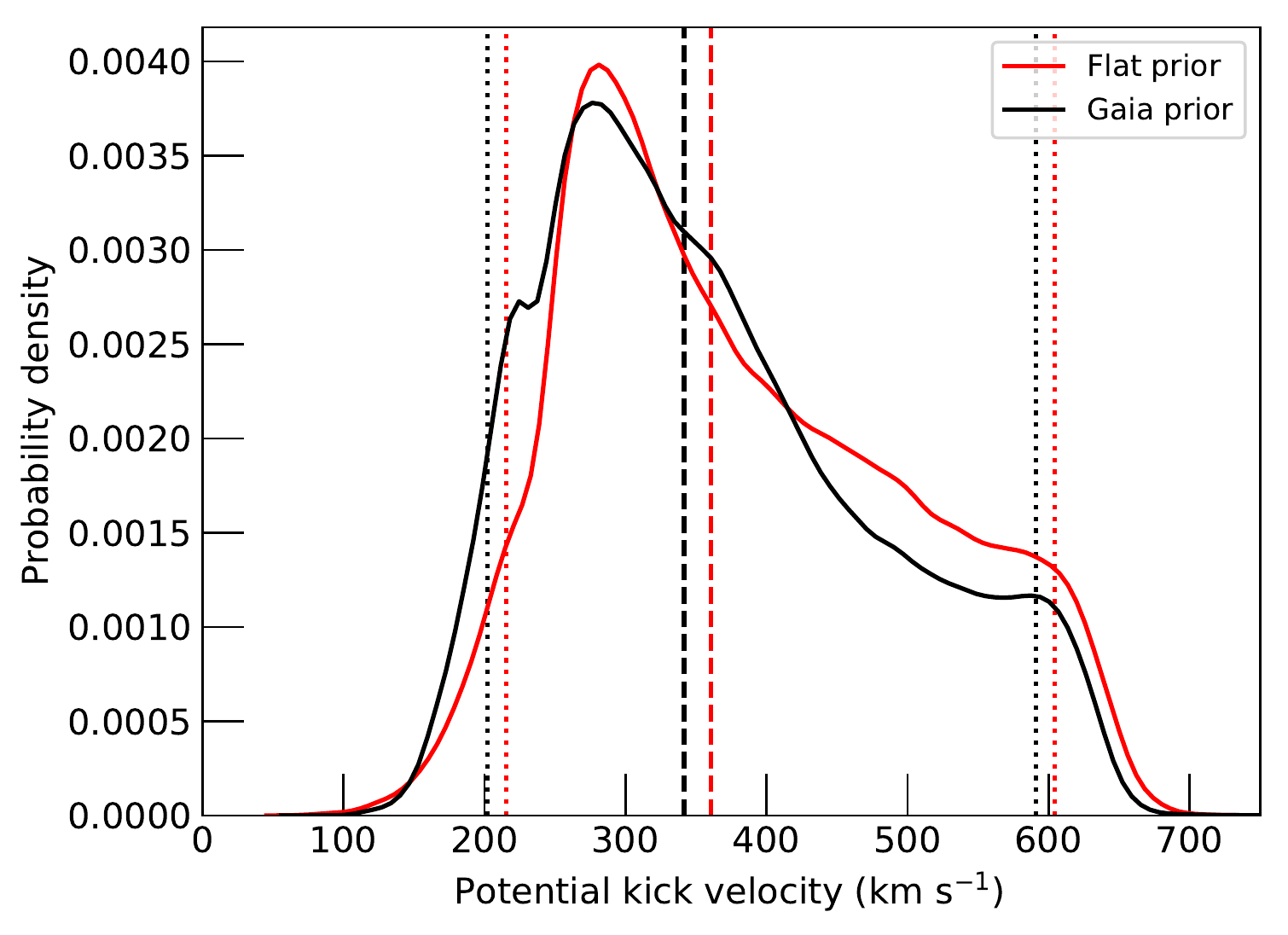}
\caption{Left panel: The posterior distance distribution of 4U\,1957+11, using the {\it Gaia} parallax constraints, along with the assumptions described in the main text about the spatial distribution of X-ray binaries. Right panel: Potential kick velocity distribution of 4U\,1957+11, using both the Gaia distance prior (black solid line) and a uniform distance distribution prior (red solid line). The kick velocity distribution in both cases does not go below 100\,km\,s$^{-1}$, suggesting that the system received a strong natal kick at birth, or formation in the halo. Dashed lines in both the panels represent the median and the dotted lines represent the 5$^{\rm{th}}$ and 95$^{\rm{th}}$ percentiles of the distribution.}
\label{peculiarvel}
\end{figure*}

\subsection{The orbital period and nature of the donor star}

Because 4U~1957+11 has never gone into quiescence, its optical flux has consistently been dominated by its accretion disk, rather than by its donor star.  An orbital period of 9.33 hours has been measured from photometric modulations \citep{Thorstensen1987}, likely due to illumination of the donor star by the accretion disk \citep{Thorstensen1987, Bayless2011}. Emission lines give a tentative mass ratio of $\approx0.3$ \citep{Longa-Pena} but neither reliable direct radial velocity curves of the donor star's absorption lines nor inclination angle estimates from the orbital data have been produced.  

The donor star may be slightly evolved.  A halo star is possible, but would have to be toward the metal rich end of the halo star distribution and to be slightly evolved; 10~Gyr old stars filling a Roche lobe at a period of 9.3 hours with [Fe/H]$\approx-1$ can still have cool enough envelopes to allow convection \citep{Pietrinferni2006}.  For donors that are significantly more metal poor, stars that fill this Roche lobe will have radiative envelopes even after 10~Gyr, suppressing magnetic braking, and hence would not have substantial mass transfer rates.  

\subsection{Astrometry and possible halo nature}
The object is detected in Gaia DR2 \citep{GaiaDR2} with a parallax measurement of 0.025$\pm$0.239 mas, and a proper motion of $-0.19\pm0.36$\,mas\,yr$^{-1}$ in RA and of $-1.94\pm0.29$\,mas\,yr$^{-1}$ in Declination. After correcting for the global zeropoint offset of -0.029\,mas in Gaia's parallax measurement \citep{Luri2018,Chan2019} the parallax measurement can be used to constrain the distance to the source with the use of an appropriate prior.  For "normal" stars, \citet{Bailer-Jones2018} provide such a prior. Because X-ray binaries typically have a larger scaling height than disk stars, we use a Milky Way-like prior \citep{Atri2019} with scale height parameters derived in \citet{Grimm2002} for X-ray binaries. We determine the posterior distance distribution to the source with a median of 7\,kpc and 5$^{\rm{th}}$ and 95$^{\rm{th}}$ percentile of 3 and 15\,kpc, respectively {(see Figure \ref{peculiarvel}, left panel)}.  We note that some additional considerations based on the X-ray properties, discussed in the next subsection, disfavor the lower end of this range.

The center of mass radial velocity, $\gamma$ of the binary, is -180$^{+30}_{-38}$ km\,s$^{-1}$ \citep{Longa-Pena}.  This comes from Bowen fluorescence measurements.  Given that this source is located at a Galactic longitude of $+51$ degrees, it is expected that its radial velocity would be positive, not negative.  As a result it must have had a substantial natal kick, or it must be a halo system. We used the proper motion, the systemic radial velocity and the distance constraints derived above to constrain the current three dimensional space velocity of the system.   The age of the system is unknown but most X-ray binaries are assumed to be born in the Galactic plane.   We use \textit{galpy} \citep{Bovy2015} and Monte Carlo simulations to integrate and sample 5000 Galactocentric orbits of the system and derive the probability distribution for the peculiar velocity of the system when it crosses the Galactic plane, called the potential kick velocity \citep[PKV,][]{Atri2019}. As can be seen in Figure \ref{peculiarvel}\,(right panel), the PKV distribution of the system has a median of 346\,km\,s$^{-1}$, with a 5$^{\rm{th}}$ and 95$^{\rm{th}}$ percentile of 203 and 594\,km\,s$^{-1}$, respectively. Due to the large uncertainty on the Gaia parallax, we also test a loosely constrained uniform distance prior of 5--25\,kpc and find that the PKV distribution is not affected considerably, with a median of 360\,km\,s$^{-1}$, and 5$^{\rm{th}}$ and 95$^{\rm{th}}$ percentiles of 215 and 602\,km\,s$^{-1}$, respectively. Thus, regardless of the distance of the source the PKV of the system is always in excess of 100\,km\,s$^{-1}$. 

The PKV distribution of the source is too large to be accommodated in a straightforward manner with just the \citet{Blaauw1961} mechanism of mass loss from a moving component of a binary.  The maximum kick in such a scenario is given in \citep{Nelemans1999} as:
\begin{equation}
    v_{sys}= 213 \left(\frac{\Delta{M}}{M_\odot}\right) \left(\frac{m}{M_\odot}\right) \left(\frac{P_{\rm re-circ}}{{\rm day}}\right)^{-1/3}\left(\frac{M_{\rm BH}+m}{M_{\odot}}\right)^{-5/3} {\rm km/sec},
\end{equation}
where $\Delta{M}$ is the mass lost during the supernova (limited to half the total system mass if the system remains bound after the supernova), $m$ is the donor mass, $M_{\rm BH}$ is the black hole mass, and $P_{\rm re-circ}$ is the period of the binary after re-circularization.  Allowing for substantial mass transfer in the system to have already taken place, 100 km\,s$^{-1}$ could be achieved, but 200 km\,s$^{-1}$ cannot be achieved without very unreasonable assumptions.  The possible range of kicks could easily be produced if there is an asymmetric natal kick \citep{Brandt1995}. This would make the system the fastest moving black hole X-ray binary known in our Galaxy.

The potential halo nature of the object is intriguing.  For this object, the strengths of the Bowen fluorescence lines are typically a factor of 2 weaker than the He~II emission lines at 4686~\AA \citep{Longa-Pena}.  In most other systems, the Bowen lines are stronger than the He~II lines.  Without careful photoionization modelling, this does not definitively indicate that the abundance ratio of helium to carbon and nitrogen is anomlously high for this system, but it does indicate that this is potentially the case and hence worth such an investigation.  This tentative evidence for a halo origin plays against the low mass of the black hole, given that low metallicity environments lead to weaker winds and more massive remnants, while the highest kick black holes are expected to be the least massive ones \citep{Belcz2002}.

\subsection{X-ray properties and the question of black hole versus neutron star accretor}

There is no direct evidence for a black hole in this system.  There is some evidence that if the system contains a black hole, the black hole is toward the low mass end of the mass spectrum \citep{NowakWilms}, possibly in the "mass gap" between 2 and 5 $M_\odot$ \citep{Ozel2010,Farr2011}.   \citet{Bayless2011} make a series of arguments, none of which they claim to be conclusive, that suggest that the system contains a neutron star.  The two most important of these arguments are (1) the lack of superhumps seen in the system, and (2) the amplitude of the optical periodicity.  It is not clear that the lack of superhumps is strongly constraining; the ratio of viscous heating to reprocessing in X-ray binaries in the optical band is much lower than in cataclysmic variables, and the strength of superhump emission is significantly weaker \citep{Haswell2001,Russell2010}.  The amplitude of the orbital periodicity is also used by \citet{Bayless2011} to argue for a relatively small mass ratio, but here, the conclusions depend strongly on the inclination angle of the system and assumptions about the size of the accretion disc relative to its circularization radius and its temperature. \citet{Longa-Pena} also find evidence for a mass ratio of $0.25-0.3$ from the ratios of the amplitudes of the radial velocity curves for the accretion disc and the donor star as inferred from emission lines of both.  All these results are consistent with a black hole of relatively low mass ($\sim4 M_\odot$), as long as one understands that the optical superhumps may be quite weak in a system where the optical emission is dominated by reprocessing.  Furthermore, using a similar approach to that of \citet{Bayless2011} with a larger data set, \citet{Gomez2015} conclude that a low mass black hole is the most likely accretor.

{ The X-rays from the source consistently exhibit low amplitude variability ($\lesssim$ 2\% rms amplitude), and show a very soft spectrum, with rare instances at the faint end of the luminosity range for the source where it show some tentative increase in the characteristic radius of the accretion disc and strength of the power law component, possibly indicating an intermediate state \citep{Nowak2012}}.  Black hole X-ray binaries show this behavior in their "soft states", which typically occur above 2\% of the Eddington luminosity \citep{Maccarone2003,Armin2019}.  Neutron stars tend to show this low amplitude variability behavior in ``banana states", which also occur in the same Eddington fraction range \citep{Maccarone2003}.  This likely gives a distance of $\approx15$~kpc for a black hole X-ray binary, and for a neutron star X-ray binary, it gives a value of $\approx 7~$kpc.  The most likely range of distances thus excludes most of the range with the slowest velocities, bolstering the case made above for a halo orbit, regardless of whether that orbit results from birth in the halo or a strong natal kick.  

Beyond this, we can also consider the likely range of accretion rates for the source based on binary stellar evolution.  Re-scaling equation (9) of \citet{KKB} (and leaving out the negligible gravitational radiation component), we can see that, for a 9.33 hour orbital period, the mass transfer rate can be expected to be:
\begin{equation}
\dot{m}_2 =4\times10^{17} \left(\frac{m_1}{3 M_\odot}\right)^{-2/3} \left(\frac{m_2}{0.93 M_\odot}\right)^{7/3}~{\rm g\,s}^{-1},
\end{equation} 
with the mass transfer due to magnetic braking.   Taking this accretion rate, we find that if the donor is a main sequence star, the X-ray luminosity will be $6\times10^{37}$ erg\,s$^{-1}$ if the source is an accreting neutron star, at about 30\% of the Eddington luminosity for a standard accreting neutron star, making it likely to show the strong variability characteristic of the Z-sources.  For an accreting black hole, it will be at about 2--4$\times10^{37}$ erg\,s$^{-1}$, perhaps a bit brighter if the black hole is rapidly rotating and hence more radiatively efficient, as has been suggested based on its hot accretion disc \citep{Nowak2012}.  If the donor star is a mildly evolved halo star, the $m_2^{7/3}$ term may cancel out the increased efficiency due to the deeper potential well for the rotating black hole.  Binary evolution thus provides another argument in favor of a black hole nature (in addition to the X-ray spectral considerations already presented in \citealt{Nowak2012} and the lack of Type~I bursts discussed above).  The sum total of these arguments is that there is reasonably good evidence for a black hole of lower mass than is typically seen in LMXBs, but that there is also strong, albeit not dynamical, evidence against a neutron star.

\section{The jet power suppression rate}
The $3$-$\sigma$ upper limit on the radio flux density from the source is 3.2 $\mu$Jy.  In \citet{2011ApJ...739L..19R}, we established a flux density limit of 11.4 $\mu$Jy, which yields a jet radio power suppression of a factor of 330 (at 20~kpc) to 810 (at 7~kpc) relative to the standard black hole X-ray binary correlation from \citet{Gallo2003}.  Retaining the same distance range in \citet{2011ApJ...739L..19R}, we find that the suppression given the 3-$\sigma$ flux upper limit is a factor of 1500-3700 relative to the radio/X-ray correlation in the hard state, with some of the additional size of the effect coming from the X-ray flux being 1.25 times as high as in those observations.  If we restrict ourselves to the upper end of the distance range, based on the arguments above related to the likely accretion rate of the source, then we end up toward the lower end of the range.

The other sources which have shown comparably extreme jet suppression are H1743-322 \citep{Coriat2011}, with a factor of about 700 and MAXI J1535-571 \citep{Russell2019} with a factor of about 3000.  We note that the presence of a "radio-quiet" track in the hard state \citep{Coriat2011} does not strongly affect this result --- these systems have steeper correlations than those in \citep{Gallo2003}, but join to the standard correlation for the brightest and faintest hard states.

\section{Discussion}

\citet{Meier2001} have worked out expected jet powers for different accretion disk models.  While numerical calculations have advanced dramatically since the work of  \citet{Meier2001}, it is still the case that
the most sophisticated treatments of general relativistic MHD work best for systems at low accretion rate, because increasing the accretion rate increases the computational time (e.g. \citealt{Liska2019}).  As a result, we still use the semi-analytic work of \citet{Meier2001} as our basic framework for interpreting the results.

\subsection{Radiatively inefficient jets: unlikely in light of Cyg~X-3}

An alternative to the scenario in which the jets have their power suppressed is one in which power is injected into the jets, but the power is not effectively dissipated, so that the ratio of radio power to kinetic power is substantially smaller than in hard states.  In most systems, this could occur for a few reasons -- Poynting flux domination \citep{Lovelace2002} or poor dissipation of the power due to the lack of variability \citep{Drappeau2017}.  The results on Cygnus X-3, where the return from the soft to hard state is where the strongest jet emission is seen \citep{Koljonen2018}, strongly argue that the jets genuinely turn down in power during the soft state.  In principle, Poynting flux could pass through the stellar wind without being dissipated, and numerical calculations specific to this problem, but outside the scope of this paper should be done to test this possibility, but one of the original motivations for considering Poynting flux jets over matter-dominated jets was to {\it increase} the radiative efficiency (\citealt{Giannios2005}).  The Cyg~X-3 results are more clearly problematic for the weak dissipation model of \citet{Drappeau2017}.  Shocks will necessarily develop against the stellar wind, whose velocity is roughly perpendicular to the jets in the inner jet region, and this will be true regardless of the level of velocity variability in the ejected material.

\subsection{A Schwarzschild black hole in 4U~1957+11: unlikely to be the case or to cause the effect}

At least according to theoretical work, jet power can be strongly
suppressed in all spectral states if a black hole has a low rotation
rate. Observational work on this topic indicates mixed results;
relatively little evidence exists for spin affecting the production of
jets in hard states (e.g. \citealt{Fender2010,Russell2013,McClintock2014,Middleton2014}), and may result from the correlation of both inferred spin and peak $L_X$ with period \citep{Wu2010}.   The spin-period correlation may have a physical origin if e.g. black holes grow dramatically after birth \citep{Fragos2015}, while the period-X-ray peak correlation has a very clear physical explanation in the larger accretion discs at longer period \citep{Wu2010}.

Several investigations of the spin of this black hole have been made, via disc continuum fitting \citep{Nowak2012,Maitra2014}.  All favor the idea that the black hole is rapidly rotating, and furthermore find that the distance must be at least 10~kpc in the context of this model (and that would require a low black hole mass of 3$M_\odot$ -- \citealt{Nowak2012}).

\subsection{Fundamentally low jet power?}

The alternative then, is that the jet power is fundamentally lower in the soft state than in the hard state.  One possibility is that the analytic work of \citet{Meier2001} is not a good approximation to reality,
and the process of accelerating the jets, and that the magnetic forces are much less effective at extracting power from soft state black holes relative to hard state black holes.  If the scalings of jet power to disk properties are correct, then this would indicate that the accretion disks in soft states are substantially thinner than the approximations used by \citet{Meier2001}, or, more likely, that the ad hoc assumptions made about the ratios of their large scale magnetic fields to their dynamo fields scale .  Additionally, the low mass of the black hole may be responsible for a factor of $\sim1.5$ reduction in the ratio of radio power to jet power\citep{Merloni2003}.  We regard this combination of effects as the most likely possibility at the present time, but we also encourage more theoretical work to determine if the jet power may be genuinely suppressed by a larger factor in the soft states than previously estimated.  

\section*{acknowledgments}

The National Radio Astronomy Observatory is a facility of the National Science Foundation operated under cooperative agreement by Associated Universities, Inc.  This research has made use of the MAXI data provided by RIKEN, JAXA and the MAXI team.  We thank Poshak Gandhi for useful discussions about the distance to 4U~1957+11.  We thank the referee, Mike Nowak, for a helpful report that has improved the quality and presentation of the results.

Data Availability Statement: The VLA data presented in this paper are available from the National Radio Astronomy Observatory archives.  The MAXI data presented in this paper are available from the RIKEN web page tabulating MAXI lightcurves.  The RXTE data in this paper are available from HEASARC.

\bibliographystyle{mnras}
\bibliography{1957_radio} 

\bsp
\label{lastpage}
\end{document}